\documentclass[12pt,onecolumn]{article} 
\usepackage{amsmath}
\usepackage{amssymb}
\usepackage{latexsym}
\usepackage{graphicx}

  \newcommand{\gea}{\textsc{Geant4}}
  \newcommand{\eu}{\textsc{Eureca}}
  \newcommand{\so}{\textsc{Sources4A}}
  \newcommand{\emp}{\textsc{Empire2.19}}
   \newcommand{\wi}{\textsc{WIMP}}
\newcommand{\an}{($\alpha$,{\itshape{n}})\hspace{1,4mm}}
 \newcommand{\f}{Fig.~}

\title{Calculation of neutron background for underground experiments}
\author{V. Tomasello$^{1,2}$ , V. A. Kudryavtsev$^{1}$ , M. Robinson$^{1}$ }
\begin{document}
\maketitle

\vspace{0.1cm}

{\center \footnotesize \it $^{1}$ Department of Physics and Astronomy, 
University of Sheffield, Hounsfield Road, Sheffield S3 7RH, UK}
{\center \footnotesize \it $^{2}$ Physikalisches Institut, Eberhard Karls 
Universit\"{a}t T\"{u}bingen, Auf der Morgenstelle 14, T\"{u}bingen D-72076, Germany}
\vspace{0.2cm}

{\tiny  Corresponding author:\\ Vito Tomasello\\ Department of Physics and Astronomy, 
University of Sheffield, Hounsfield Road, Sheffield S3 7RH, UK \\
 Tel.: +44 114 22 23553; fax: +44 114 22 23555. \\
  E-mail: v.tomasello@sheffield.ac.uk}

\begin{center}

\baselineskip=24pt

{\Large \bf}

\baselineskip=18pt

\vspace{0.1cm}

\vspace{0.2cm}
\begin{abstract}

New generation dark matter experiments aim at exploring the $10^{-9}-10^{-10}$ pb 
cross-section region for the \wi-nucleon scalar interactions. 
Neutrons produced in the detector components are one of the main factors 
that can limit detector sensitivity. Estimation of the background from this source
then becomes a crucial task for designing future large-scale detectors. 
Energy spectra and production rates for neutrons coming from radioactive contamination 
are required for all materials in and around the detector. 
In order to estimate neutron yields and spectra, 
the cross-sections of \an reactions and probabilities of transitions to different excited
states should be known. 
Cross-sections and transition probabilities have been
calculated using \emp~for several isotopes, and for some isotopes, a comparison with 
the experimental data is shown. The results have been used to calculate the neutron 
spectra from materials using the code \so. 
Neutron background event rates from some detector components in a hypothetical 
dark matter detector based on Ge crystals have been
estimated. Some requirements for the radiopurity of the materials have been deduced from
the results of these simulations.

\end{abstract}

\end{center}

\vspace{0.2cm}


\section{Introduction}\label{intro}

There are many experiments, existing or planned, located deep underground. 
Such experiments are designed to detect particles coming from astrophysical sources 
like the \wi s --
Weakly Interacting Massive Particles, or astrophysical neutrinos, or to search for weak
processes, for instance neutrinoless double beta decay, predicted by new theories beyond 
the Standard Model.
Underground physics deals with these extremely rare phenomena, 
which are very difficult to distinguish from other more ordinary signals
that come from cosmic rays and natural radioactivity.
The best way to screen from cosmic rays is to carry out experiments deep underground, 
because in this way the surrounding  layers of rock reduce their flux  by several orders of 
magnitude ($10^{5}-10^{7}$ with respect 
to the surface). Nevertheless, many radiation backgrounds are still present, primarily
originating from radioactivity.
The radiation background is crucial for the sensitivity of the aforementioned experiments to 
expected events. Knowledge of background is therefore essential for designing 
detectors (in particular for dark matter searches, such as \cite{eureca07,Bruch:2007zz, Aprile:2008zz,Brunetti:2004cf,ardm,Muna:2007zz,lux,smith}), 
their shielding and active veto systems, for 
finding ways of suppressing/rejecting this background and for 
calculating detector sensitivity.
Nowdays in dark matter searches large-scale detectors are required to 
increase sensitivity to \wi~ 
interactions, since the rate of events is predicted to be very low 
(between 1 and 10$^{-5}$ event/kg/day).

One of the main tools for background studies is the use of extensive simulation work. 
This can help with studying background suppression or rejection strategies, 
and investigating requirements on the depth, the amount of active/passive shielding, 
the purity of materials, the veto efficiency, etc. for a given experiment.

There are two main sources of background: 
local radioactivity and cosmic-ray muons. In this work we 
will focus only on the radioactivity component.

Radioactive decays result in the production of four types of particles: alphas, electrons/positrons, gamma-rays (X-rays) and neutrons. Alphas and electrons cannot travel far from the source: 
they quickly lose energy due to ionization and stop within a few microns (alphas) or 
millimetres (electrons/positrons) from the source. They can generate visible effects in a 
dark matter detector only if they are produced either within the fiducial volume of the 
detector or very close to it. 
Some of the background particles due to natural radioactivity (mainly gamma-rays)  can be 
suppressed by means of passive shielding or rejected by specific identification tools, depending 
on the experimental design (\emph{e.g.} electron recoil discrimination for dark matter 
experiments and
single-site selection for neutrinoless $\beta\beta$ decay experiments). Neutrons
are more difficult to suppress. Indeed, since single neutron interactions 
can mimic nuclear recoils produced by WIMPs, their background is crucial in 
direct dark matter searches. 

Neutrons are produced by local radioactivity via spontaneous fission and \an reactions.
The \an reactions are initiated by alpha-particles from radioactive decays of
uranium and thorium present in minute quantities in rock and detector components. 
Uranium and thorium decay chains are usually assumed to be in secular equilibrium.
A significant contribution to the neutron flux from spontaneous fission is present only for $^{238}$U.

To protect a detector from background radiation from rock and laboratory walls, 
shielding made of high-Z
materials (against gamma-rays) and low-A materials (against neutrons) can be 
installed around the detector 
\cite{carson04,araujo05,lemrani06,Stewart:2007ir,Akerib:2006ks,Wulandari:2004bj,Fiorucci:2006dx}.
However, the shielding itself and detector components
also contain the traces of radioisotopes, becoming sources of background themselves.
Materials with radioactive content as low as possible, such as copper for instance,
should be used in detector construction. 
There are also mechanical constraints in choosing the material for the
support structure of the detector. Hence a reasonable compromise between mechanical
and radioactivity requirements should be found and the calculation of the background rate
from detector components is a key factor for any experiment. 

Radioactive background transport through the experimental setup is 
usually carried out using multipurpose particle transport codes, such as 
\gea \cite{geant4}, {\textsc {Mcnpx}} \cite{Mcnpx} and {\textsc {Fluka }}\cite{fluka}.
All these codes are platforms for the 
simulation of the passage of particles through matter, using Monte Carlo methods. 
The main inputs for simulations are the energy spectra 
and production rates of particles of interest. 
These quantities are not easily obtainable, especially in the case of neutrons. 
Accurate calculation of neutron yields and spectra, for all materials used in dark matter 
searches, is crucial for designing experiments and predicting their sensitivities.

The aim of  this paper is to present and describe the procedure to calculate neutron yields and
spectra from radioactivity for most of materials relevant for dark matter searches. 
Results from \gea~ simulations of neutron background 
will also be shown for the first time for a hypothetical
large-scale cryogenic detector based on Ge crystals. 
The simulations are the starting point for extensive work to be carried out for
planning and designing the forthcoming dark matter experiments, such as
\eu~(European Underground Rare Event Calorimeter Array) \cite{eureca07}. \eu~
is proposed as a multitarget experiment (scintillators and Ge crystals) but the current 
work considers only Ge as a target. The aim of
future large-scale experiments is to explore scalar \wi-proton cross-sections in 
the $10^{-9}-10^{-10}$ pb region with a target mass of up to one tonne.

We start in Section \ref{empire} by calculating \an cross-sections using the code 
\emp~\cite{empire:2007}.  
Calculation of neutron yields and energy spectra 
from spontaneous fission and \an reactions using the code \so~\cite{ss} is discussed in 
Section \ref{spectra}. 
These spectra are the essential input to \gea~ and other particle transport codes
for the neutron propagation and detection. 
We show simulation results for neutrons from radioactivity
and their effect on the performance of a hypothetical Ge dark matter detector 
similar to \eu.

\section{\an cross-sections}\label{empire}

A typical background simulation for a dark matter experiment deals with the production and 
propagation of neutrons and gamma-rays through  the experimental setup. 
The rate of production along with the energy spectra of these particles are 
required.
Spectra of gamma-rays are available from databases, but this is not the case for neutrons. 
Although neutron yields can easily be estimated (either experimentally or by calculation),
if the \an cross section is known as function of energy, the energy spectra of neutrons
are difficult to measure or calculate. The measurements of neutron spectra
are not straightforward since
neutrons are neutral particles.
Unlike gamma-rays, the neutron spectra depend on the material, 
so they have to be obtained for all materials relevant to dark matter searches. 

The calculation of the neutron spectra requires as input the cross-sections of \an reactions 
and the transition probabilities to different excited states (branching ratios). 
As for many isotopes there are no available experimental data, the calculation of these
cross-sections is necessary. Even when the data are available, they do not usually include
probabilities of transitions to excited states (branching ratios) and therefore cannot be 
used to generate neutron spectra. For the calculation of \an cross-sections and 
branching ratios we used the code \emp. 
We extended the calculations carried out in Ref.~\cite{lemrani06}, 
by adding more materials and making other improvements.

\emp~ is composed of several modules of nuclear reaction codes, 
including various nuclear models. It is designed for calculations of cross-sections and other
parameters
over a broad range of energies and incident particles, using any nucleon or 
heavy ion as the projectile. The code accounts for the major nuclear reaction 
mechanisms (see Ref.~\cite{empire:2007} for details). A comprehensive 
library of input parameters cover nuclear masses, optical model parameters, 
ground state deformations, discrete level decay schemes, level densities, 
fission barriers, moments of inertia, and $\gamma$-ray strength functions.

Input to \emp~ consists of three parts. The first is mandatory and contains basic 
data necessary to specify the physics case (incident energy, target, projectile, 
output particle etc.). This is followed by optional inputs
which allow modifications to the default model parameters. 
In our simulations we used the optical model with Gilber-Cameron 
level densities adjusted to experimental parameters and to discrete levels.
Finally there is 
the list of incident energies for the projectile ($\alpha$-particle in our case). 

We realized that the results are sensitive to the values of incident  
alpha energies present in the input file of \emp, for which we wanted to 
calculate the cross-sections and branching ratios. 
Choosing a smaller step in energy (bigger number of input energies) 
results in an improvement in the accuracy. We reduced the step in energy 
until the variation became insignificant. The step in alpha energy that was used
in this work, was 0.1 MeV. \emp~ is also able to calculate cross-sections 
for the ground and all excited states. The branching ratios were derived from these
cross-sections.

Hence, we achieved significant improvements both for cross-section and branching ratio data
over previous calculations \cite{carson04,lemrani06}.
The number of energy points for which the cross-sections and branching ratios were
calculated was increased considerably. We also managed to 
calculate cross-sections for all possible excited states 
(in the previous work of Ref.~\cite{lemrani06} some of them were missing).

\emp~ also accounts for the continuum spectrum of gamma-rays after the 
nuclear transmutation. 
We cannot define an energy value for the continuum, 
but it is a required input for calculation of neutron yields and spectra. 
Following the suggestion of the \textsc{Endf} format  
user's guide \cite{endf} for this case, the sum of the energy threshold of the reaction 
and the threshold for the continuum level was set as
an energy value for the continuum.

The code \so~ that was used to calculate neutron yields and spectra as described 
in Section \ref{spectra}, was changed accordingly to store an increased amount of data 
in the library.  

\begin{figure} 
\begin{center}
$\begin{array}{c}
  \includegraphics[height=9cm, width=9cm]{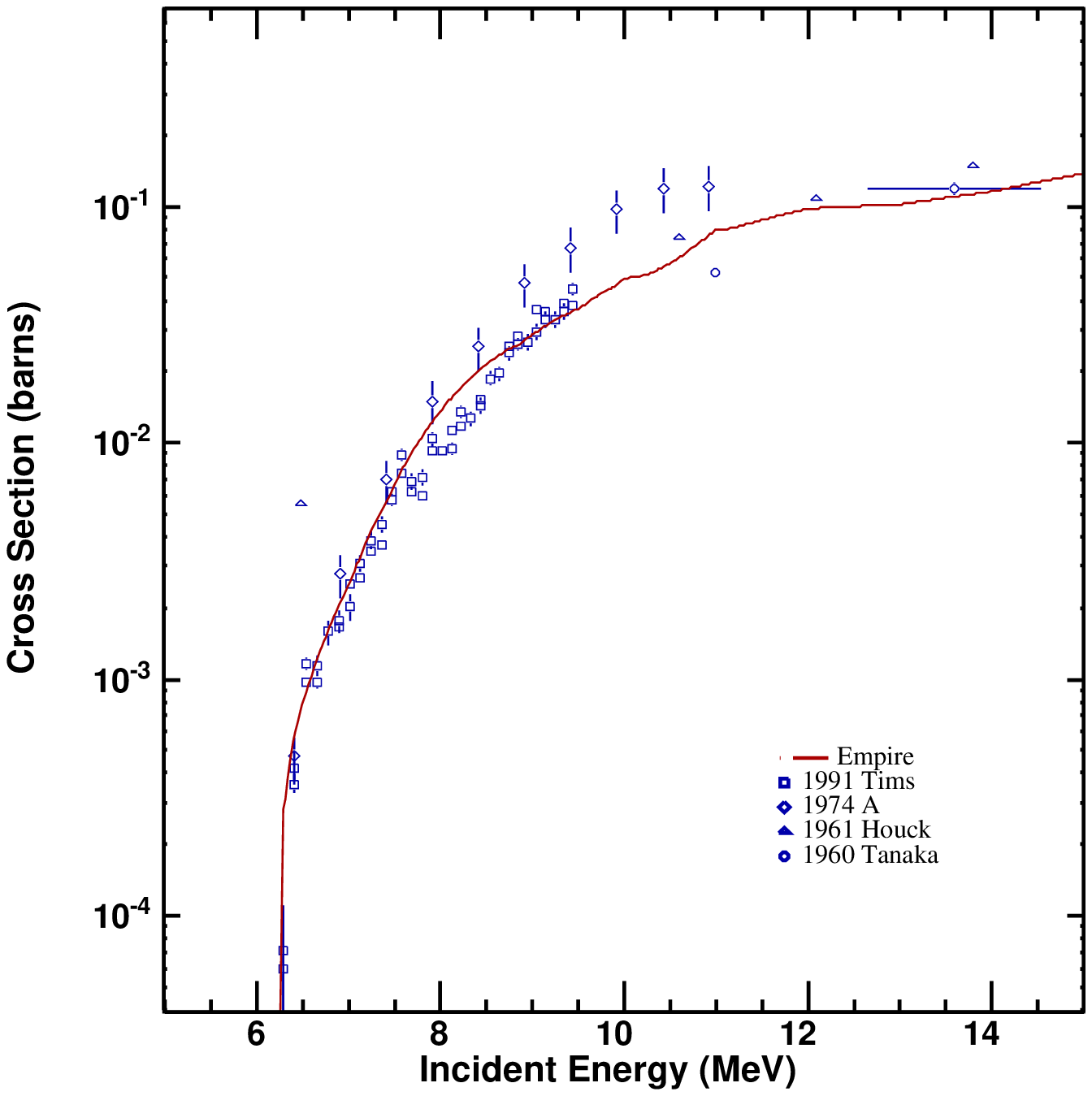} \\
(a)\\
 \includegraphics[height=9cm, width=9cm]{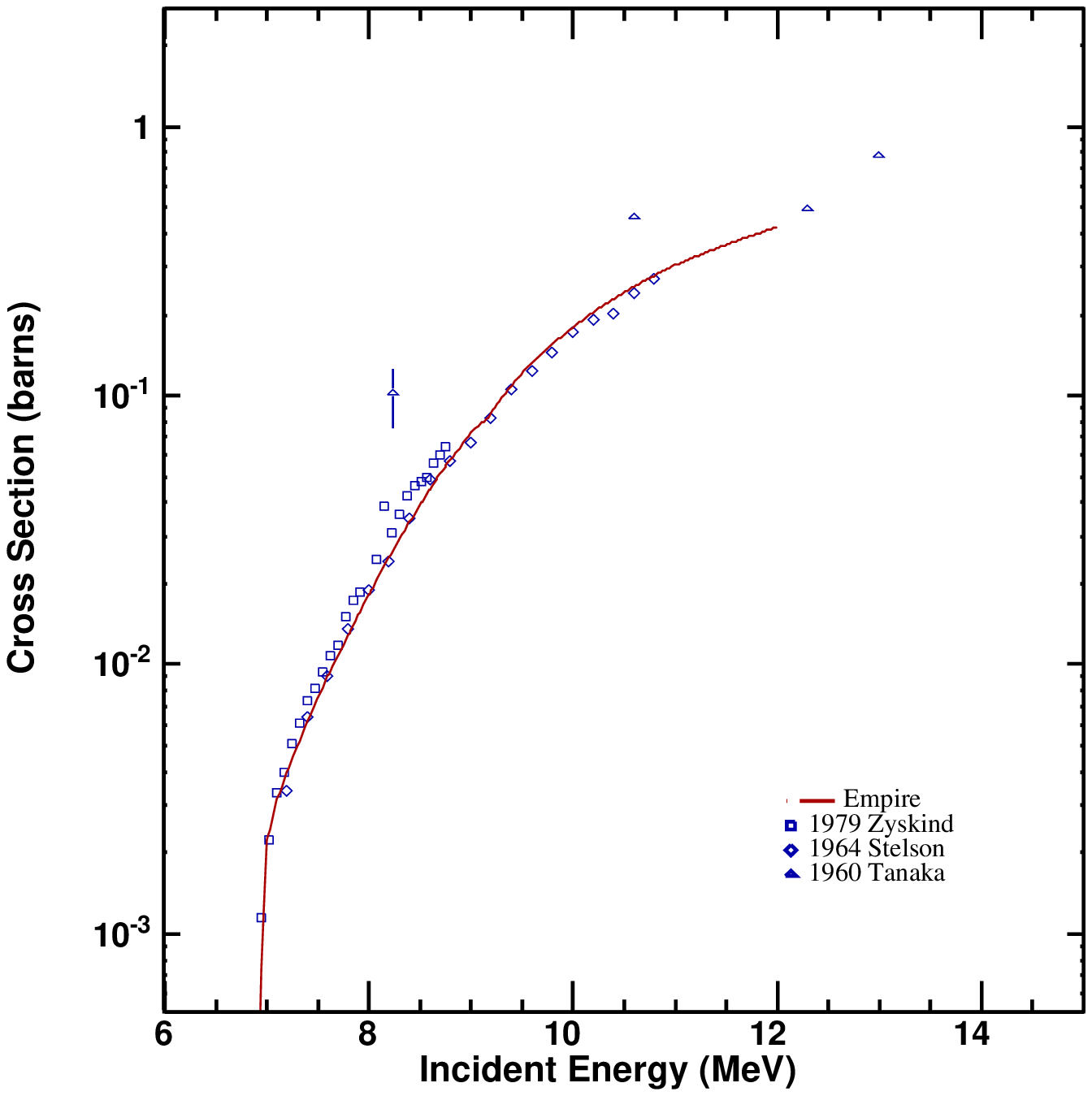}\\
 (b)\\
\end{array}$
\caption{\an cross-sections as a function of alpha energy. Results obtained with \emp~ 
(solid curve) are compared with experimental data: (a) $^{54}$Fe and (b) $^{62}$Ni. 
For the experimental data the \textsc{Exfor}~\cite{exfor} library was used.}
\label{fig-empire}
\end{center}
\end{figure}

To prove the reliability of \emp~code, we carried out calculations of \an cross-sections
for some isotopes for which experimental data are available. Comparison between 
\emp~calculations and experimental data is shown in Fig.~\ref{fig-empire} 
for $^{54}$Fe and $^{62}$Ni. There is a good agreement between data and simulations, 
especially for energies below 10 MeV, 
important for alphas from uranium and thorium decay chains. 
This agreement gives confidence in the \emp~results for others isotopes, 
for which direct comparison is not possible because of the absence of experimental data.

\section{Neutron yields and spectra}\label{spectra}

The evaluation of neutron yields and spectra can be carried out using the code 
\so.
The code calculates neutron yields and spectra from \an reactions, 
spontaneous fission, and delayed neutron emission due to the decay of radionuclides.
Its library contains all alpha emission lines from most radioactive isotopes.
The code takes into account the energy losses of alphas, cross-sections of
\an reactions and the probabilities of nuclear transition to different excited states. 
We used an option of thick target neutron yield that allows calculation 
of neutron yields and spectra under the assumption that the size of a radioactive sample
exceeds significantly the range of alphas.

Despite the tested reliability (see \cite{ss,kudAip2005}), the way in which it was 
 conceived presented two limitations. The first comes from the fact that  the original 
code provided a treatment of \an reactions only up to 6.5 MeV. 
This restricted significantly the reliability of the results, because the cross-section 
of \an reactions rises with energy and the average neutron energy also increases 
with the parent alpha energy. Therefore a modified version of \so~\cite{carson04} 
was used in this work. In this version, alpha energies up to 10 MeV are allowed.
Another problem with original \so~was that the cross-sections and branching ratios 
for transitions to the excited states for many isotopes relevant to dark matter searches
were missing from the code libraries.

As described in Section \ref{empire}, cross-sections and branching 
ratios for several isotopes were calculated 
with \emp~and inserted into the libraries of \so. The code was then used to calculate
neutron yields and spectra from decay chains of U and Th 
for many materials present in rocks and possible detector
components. If needed, other materials can easily be added.
Usually a concentration of 1 ppb of U or 1 ppb of Th in all calculations was assumed, 
which can easily be scaled to the actual concentrations.
In all calculations so far the U and Th decay chains were assumed to be in secular equilibrium.

The most critical materials to be considered as neutron sources, are mainly: 
(1) rock and concrete in the walls of the underground laboratories, that dominate the 
total neutron flux before shielding,  
and (2) materials that compose the internal detector parts (\emph{e.g.} stainless steel, copper, 
photomultiplier tubes etc.), that become important when the external flux is attenuated 
by the shielding.
Following these criteria, neutron spectra shown in this section are: copper, stainless steel,
the rock around the Modane Underground Laboratory (France) and some 
typical materials (teflon and polyethylene) present among the components of 
experimental setups. 

\begin{figure} 
\begin{center}
$\begin{array}{c}
\includegraphics[width=0.6\textwidth]{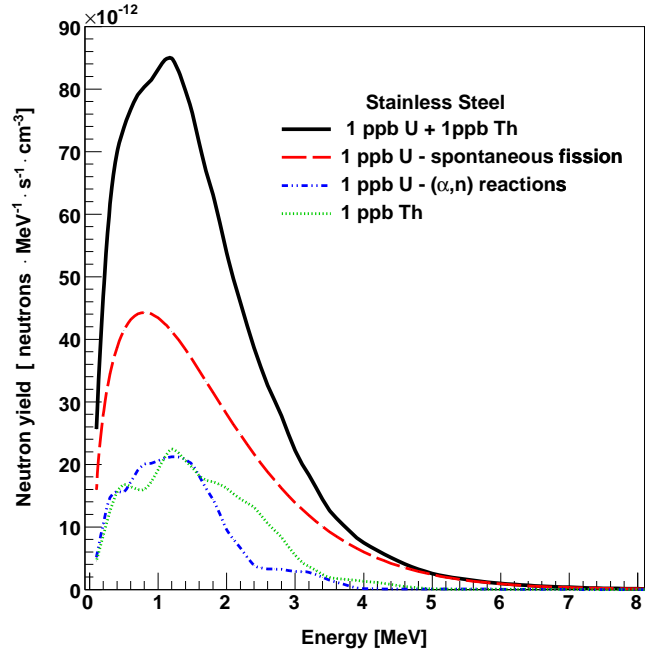} \\
(a)\\
\includegraphics[width=0.6\textwidth]{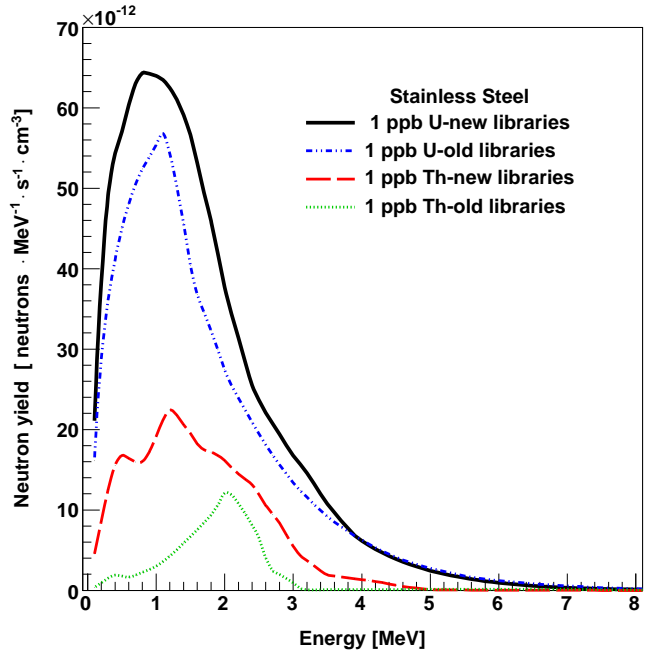}\\
(b)\\
\end{array}$
\caption{(a) Neutron spectra from stainless steel with 1 ppb of U and 1 ppb of Th. 
(b) The spectra from present work
are compared  to those calculated using different \so~libraries. 
The \lq\lq new'' library refers to our present results, the \lq\lq old'' library refers to 
Ref.~\cite{carson04}.
The neutron yields for 1 ppb of U and 1 ppb of Th are 
1.47$\times10^{-9}$ neutrons$\cdot$s$^{-1}\cdot$cm$^{-3}$ (U) and
4.74$\times 10^{-10}$ neutrons$\cdot$s$^{-1}\cdot$cm$^{-3}$ (Th) for \lq\lq new'' library, and
1.19$\times 10^{-9}$  neutrons$\cdot$s$^{-1}\cdot$cm$^{-3}$ (U) and 
1.56$\times 10^{-10}$ neutrons$\cdot$s$^{-1}\cdot$cm$^{-3}$ (Th) for the old ones.
The mean neutron energies are: 
$<E_n>$=1.58 MeV (U), $<E_n>$=1.56 MeV (Th) for the new libraries and
$<E_n>$=1.66 MeV (U), $<E_n>$= 1.92 MeV (Th) for the old ones.
See text for details.}

\label{steel}
\end{center}
\end{figure}

The spectra from stainless steel are presented in Fig.~\ref{steel}a, showing
a significant contribution from spontaneous fission from $^{238}$U. 
Fig.~\ref{steel}b shows comparison of our calculations for stainless steel using 1 ppb of U 
and Th with a previous work \cite{carson04}. 
In Ref. \cite{carson04} the neutron yields were also calculated with \so, but 
the library contained a restricted set of isotopes and transitions for many isotopes
were assumed to the ground
state only. For instance, cross-sections for all Fe isotopes were assumed to be the same
as the measured cross-section for $^{54}$Fe. Fig.~\ref{steel}b shows that these
assumptions led to the overestimation of the neutron energies and underestimation of the
neutron flux.

\begin{figure}
\resizebox{0.9\textwidth}{!}{%
   \includegraphics{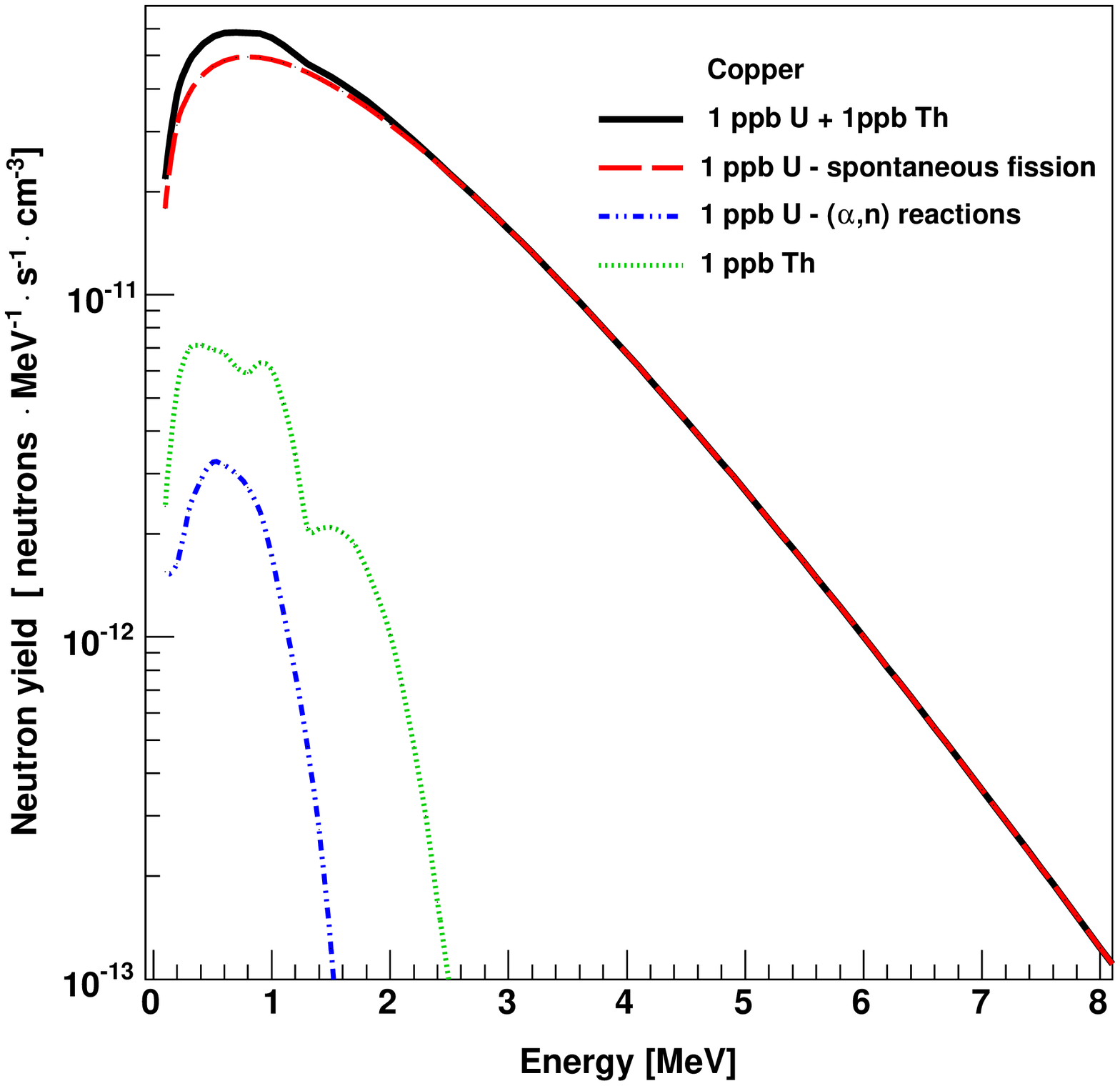}
   }
    \caption{Neutron spectra from copper assuming 1 ppb (both  U and Th) of radioactive contamination.
Different contributions are shown along  with the sum spectrum (solid curve).}
  \label{copper}
\end{figure}

The neutron energy spectra from U and Th in copper are shown in \f\ref{copper}. 
A logarithmic scale on y-axis is chosen here to make visible the small contribution from 
\an reactions. The neutron yield from \an reactions is suppressed in copper
due to the high energy threshold which is about 7~MeV 
(due to high-Z and Coulomb barrier). In this case
the main contribution comes from spontaneous fission of $^{238}$U. Hence, for a given U/Th concentration,  copper has an advantage over other materials containing lower-Z 
isotopes (for instance stainless steel).

\begin{figure}
\resizebox{0.9\textwidth}{!}{%
   \includegraphics{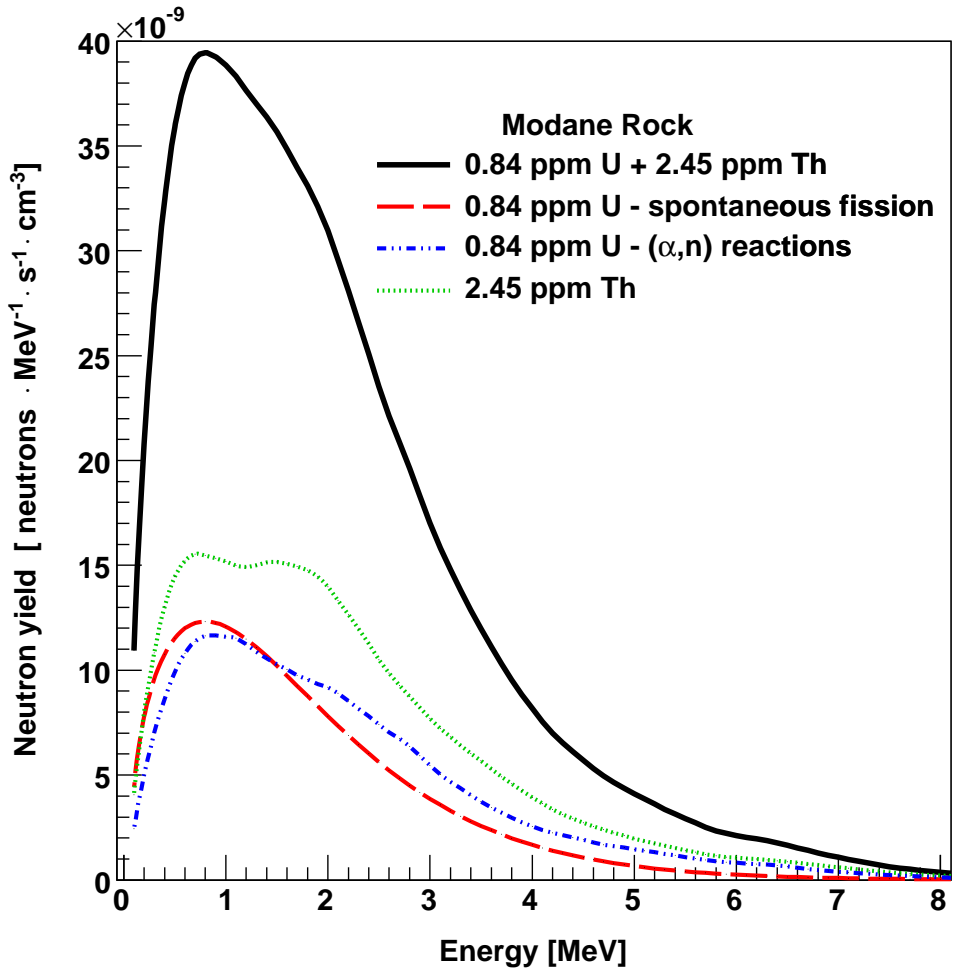}
   }
    \caption{Neutron energy spectra from uranium and thorium in the
    rock around the Modane Underground Laboratory. Rock contamination is taken 
    from \cite{chazal98}. Different contributions are shown along  with the sum 
    spectrum (solid curve).}
  \label{modane}
\end{figure}

\f\ref{modane} shows the neutron energy spectrum at production 
in rock around the Modane Underground Laboratory. It is similar to that 
presented in Ref.~\cite{lemrani06}, but has been obtained with
the aforementioned improvements to the \an cross-section calculations. 
The rock composition and U/Th concentrations were taken from  Ref.~\cite{chazal98}.  
Separate contributions from \an reactions in the uranium decay chain, thorium chain 
and $^{238}$U spontaneous fission are shown. 
The spectrum of neutrons from $^{238}$U spontaneous fission is described by the Watt 
function~\cite{watt}, with a peak energy of about 0.8~MeV and a mean energy of 
about 1.7~MeV.
As the spectra of neutrons from \an reactions are not much harder, this results in 
an overall energy spectrum with mean energy of about 1.9~MeV. 
In general, simulations with \so, taking into account proper branching ratios for transitions 
to excited states, give softer neutron spectra than reported in earlier simulations and 
measurements (see, for instance, Refs.~\cite{chazal98,arneodo99}).

\begin{figure}
\resizebox{0.9\textwidth}{!}{%
   \includegraphics{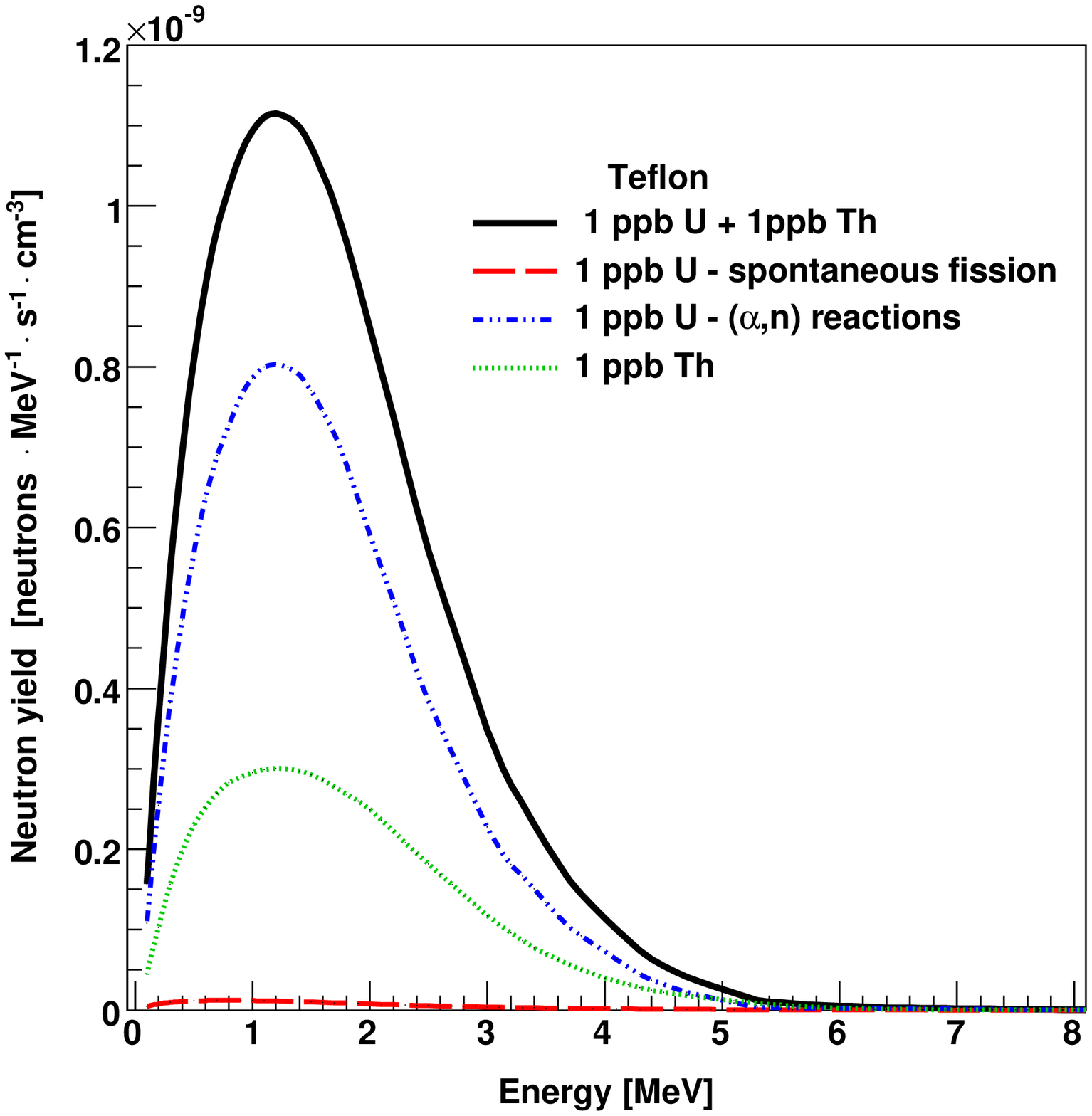}
   }
    \caption{Neutron spectra from teflon assuming  1 ppb (both  U and Th) of radioactive 
    contamination. Different contributions are shown along with the sum spectrum (solid curve).}
  \label{teflon}
\end{figure}

\begin{figure}
\resizebox{0.9\textwidth}{!}{%
   \includegraphics{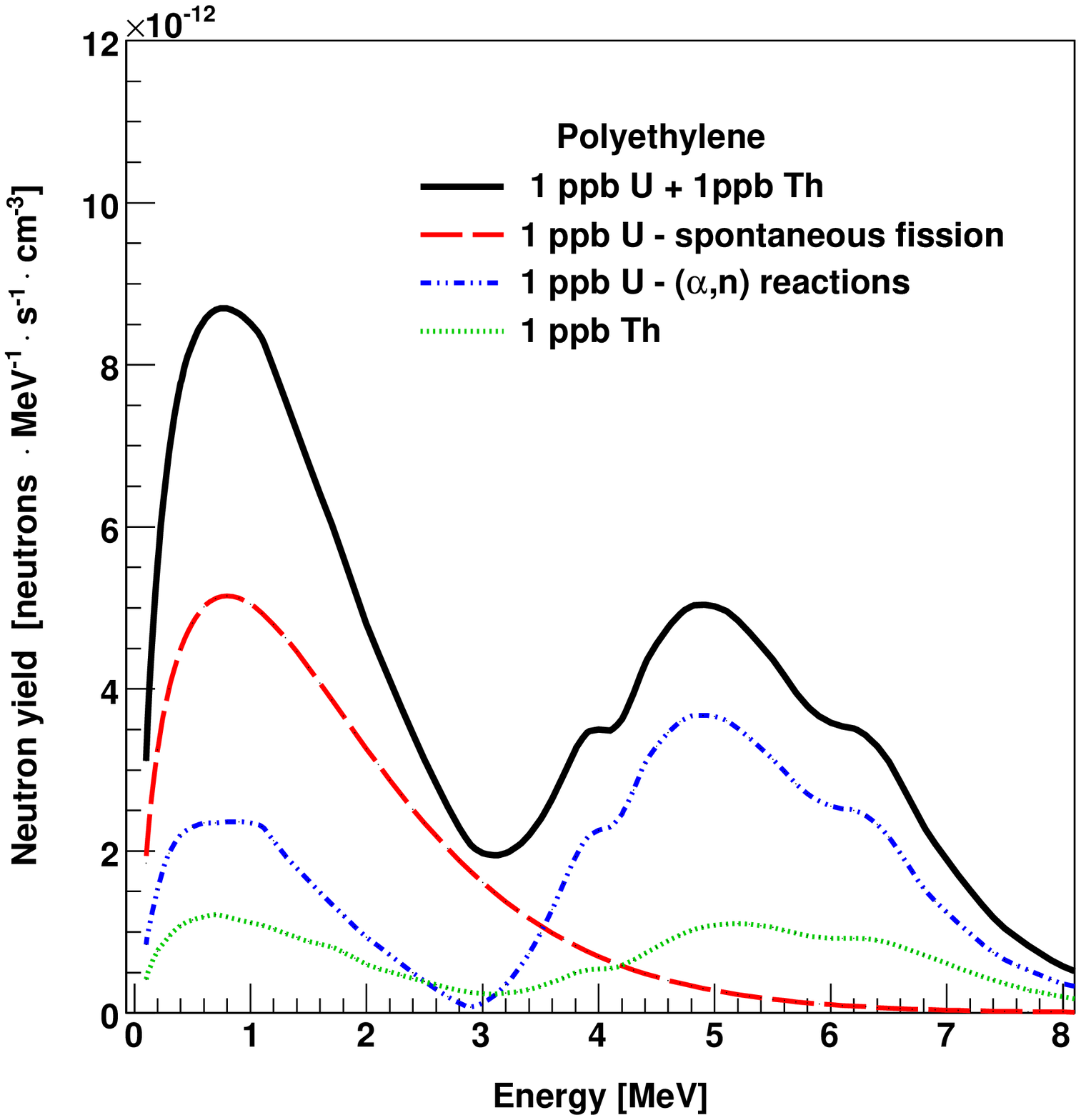}
   }
    \caption{Neutron spectra from polyethylene assuming 1 ppb (both  U and Th) 
    of radioactive contamination. Different contributions are shown along with the 
    sum spectrum (solid curve). }
  \label{polyethylene}
\end{figure}

Fig.~\ref{teflon} and Fig.~\ref{polyethylene} show neutron energy spectra from U and Th in teflon and
polyethylene. In both cases spontaneous fission is not the main contributor.
Its contribution is practically negligible for teflon. These two materials are often
used in small amounts nearby the detectors. The background neutron flux
from materials close to the sensitive volume of the detector is difficult to shield or suppress
by active rejection techniques. It is, therefore, crucial to estimate the contribution of these materials 
to the background event rate since they can limit the sensitivity of the experiments.

\section{Neutron background from detector components}

In this Section we present some results of neutron propagation and detection in 
a particle dark matter detector based on Ge. We used neutron yields and energy spectra,
as calculated using the modified version of \so~ with extended libraries, as input to 
the simulations. These  simulations form a part of a preliminary study to investigate 
radioactive background for future large-scale cryogenic dark matter detectors.

Dark matter detectors are looking for low energy depositions (keV energies) from
nuclear recoils due to \wi-nucleus interactions. 
Gamma-ray flux from radioactivity usually exceeds by several orders of magnitude 
the neutron flux. However, gamma-rays generate electron recoils in dark matter 
detectors that can relatively easily be discriminated from nuclear recoils expected 
from \wi s.

Neutrons are   more likely to cause problems   than gamma-rays since they can 
elastically scatter off
nuclei producing nuclear recoils in a similar way to  \wi s.

The simulations described here were mainly intended to study:
\begin{enumerate}
 \item The background event rate due to neutrons from copper -- the most massive 
material in the cryogenic detector construction.
\item  The change in background rate due to neutrons if stainless steel is used in 
addition to copper.
\end{enumerate}

\begin{figure} 
\begin{center}
\includegraphics[width=0.6\textwidth]{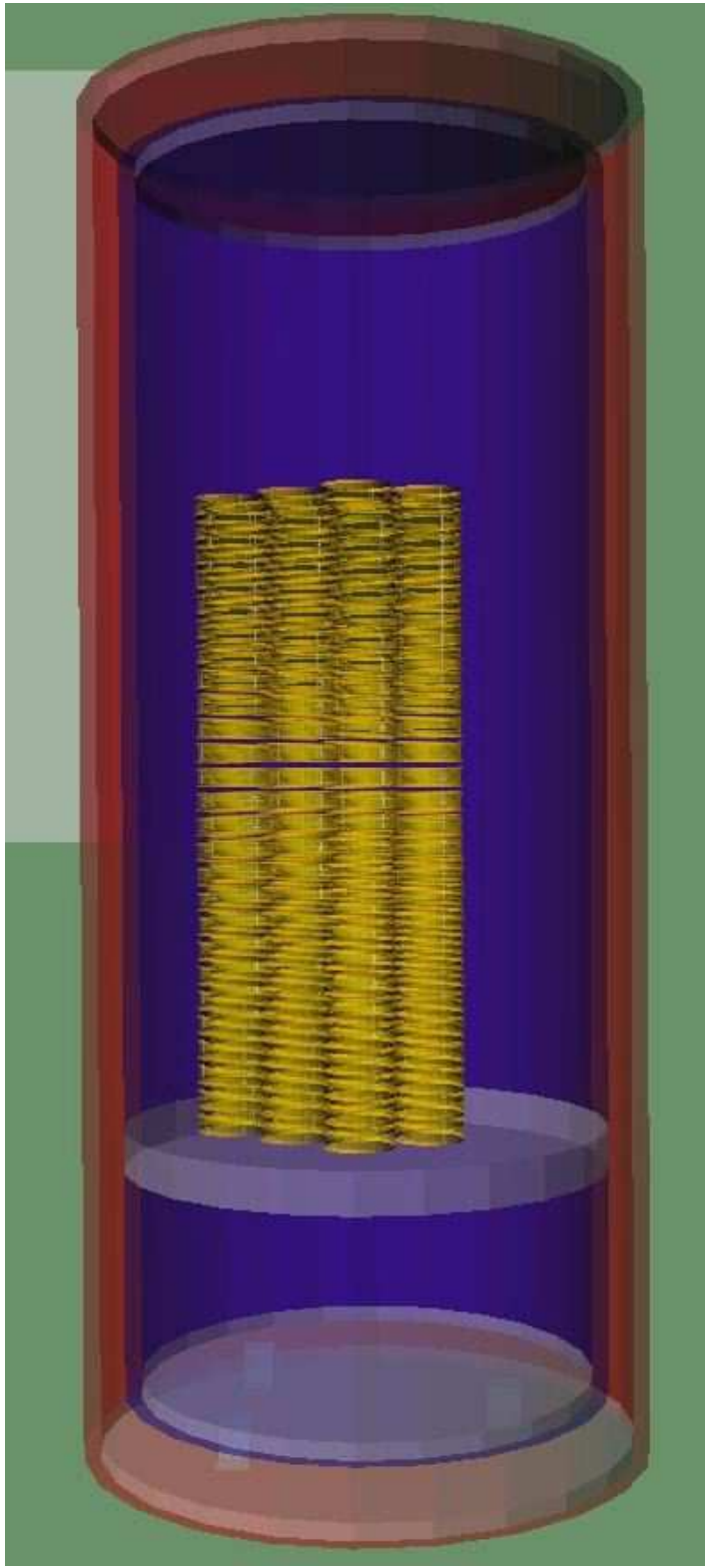}
\caption{Schematic view of copper vessels and Ge crystals from \gea.}
\label{tower}
  \end{center}
\end{figure}

The simulations were carried out using \gea.
\f\ref{tower} shows the detector  consisting of two copper vessels (inner vessel with 
thickness of 0.5~cm 
and a mass of 139 kg, and an outer vessel with a thickness of 0.5~cm and a mass of 
181~kg) 
containing 103.68~kg of Ge supported by a copper plate with a thickness of 
1~cm and a mass of 22~kg.
Crystals are arranged in 27 floors, each containing 12 crystals with a mass of 320~g 
each. Crystals are similar to those described in Ref.~\cite{geDet}.
Although we used a simplified design of a dark matter detector based on Ge target,
it enabled us to estimate the background rates due to neutrons 
from the most massive detector components.

Neutrons were generated evenly within two copper vessels and a copper plate 
below the Ge crystals. Their directions were sampled isotropically. 
Neutrons were transported to the crystals using \gea~  toolkit 
and their energy depositions in Ge were
recorded. Both elastic and inelastic scattering of neutrons were taken into account.
As a result recorded energy depositions were due to nuclear recoils and/or 
gamma-rays/electrons produced via neutron inelastic scattering in crystals or their surroundings.
In further analysis we considered only events with energy deposition in a single crystal
more than 10 keV. This energy corresponds approximately to the expected threshold
of high-sensitivity detectors.

\begin{figure}[htbp]
\begin{center}
\includegraphics[width=10cm]{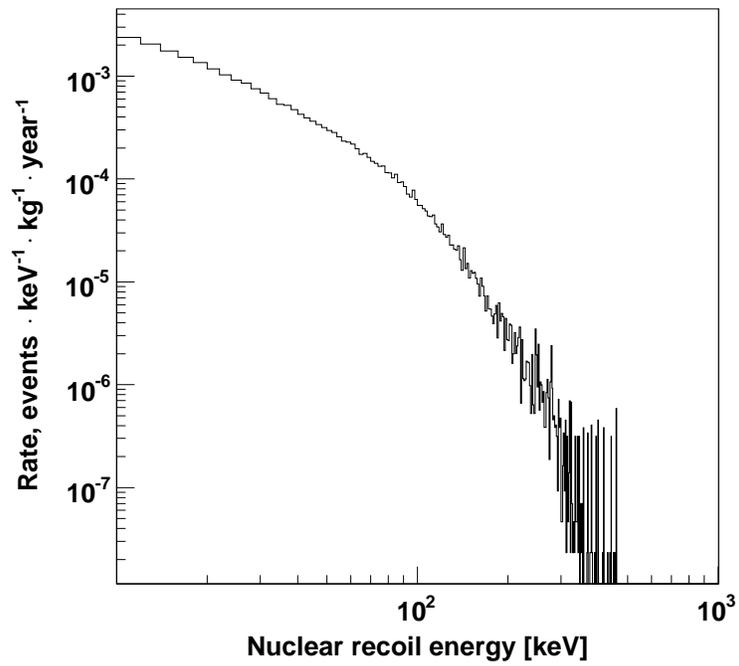}
\caption {Energy spectrum of nuclear recoils in 103.68 kg of Ge from 1 ppb of uranium 
and thorium
in the copper vessels and plate. 
The energy threshold of 10 keV was assumed to select events
(see text for details).}
\label{fig-nspec}
\end{center}
\end{figure}

\f\ref{fig-nspec} shows an energy spectrum of single nuclear recoils above 10 keV in Ge
from 1 ppb of uranium and thorium in the two vessels
and the copper plate supporting the crystals. 
To plot this spectrum the 
following selections have been made: (i) only single nuclear recoils have been selected, i.e.
if a neutron scattered more than once producing two or more nuclear recoils in two or more
different crystals with energies more than 10 keV each, the event was rejected; 
(ii) if two scatters happened in one crystal, they were considered as a single nuclear recoil, 
i.e. no position sensitivity was assumed within a single crystal; 
(iii) if an energy deposition in a crystal was less than 10 keV, it was neglected, i.e. an energy
threshold for a deposition in a crystal was set to 10 keV; (iv) if an energy deposition was due to
an electron recoil either alone or in combination with a nuclear recoil, it was neglected, i.e.
a perfect discrimination was assumed for gamma-ray induced events. The spectrum shown in 
\f\ref{fig-nspec} was obtained assuming 1 ppb concentrations of U and Th in copper that
is probably 100 times higher than could be achieved for ultra-low radioactivity copper.
 
\begin{figure}[htbp]
\begin{center}
\includegraphics[width=10cm]{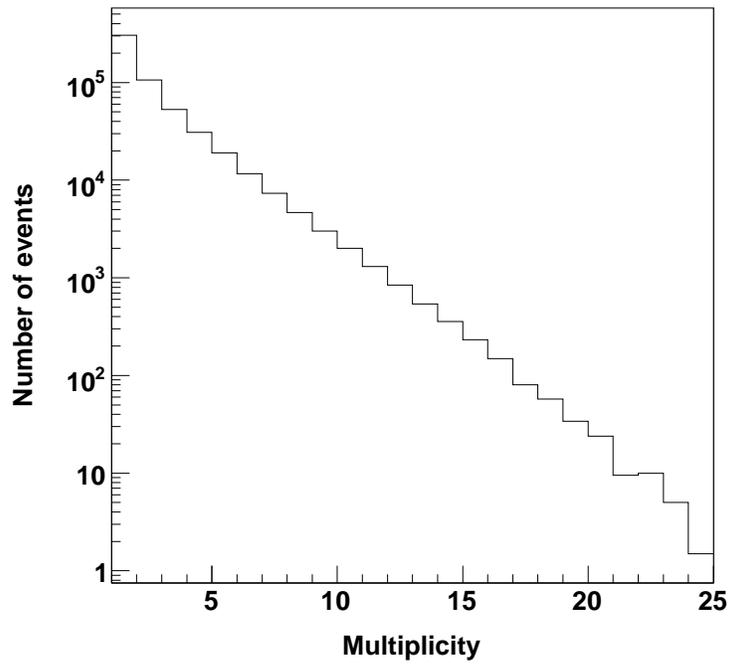}
\caption {Multiplicity distribution of energy depositions due to nuclear recoils
in Ge crystals from 1 ppb of uranium and thorium
in the copper vessels and plate. Energy depositions from nuclear and electron 
recoils are included
but at least one nuclear recoil above 10 keV is required.
The energy threshold of 10 keV was assumed to select events
(see text for details).}
\label{fig-nmult}
\end{center}
\end{figure}

\f\ref{fig-nmult} shows the multiplicity distribution of events that include nuclear recoils.
Neutrons were produced by the uranium and thorium decay chains (1 ppb of U and Th) in 
the copper vessels and supporting plate.
A single energy deposition (multiplicity equal to one) was 
defined in the same way as above. If more than one \lq\lq energy deposition'' due to
either nuclear or electron recoils occurred in 
different crystals, then an appropriate multiplicity was attributed to this event. Energy
depositions from electron recoils (due to neutron inelastic scattering) were included
in the plot. If an energy deposition in a crystal was less than 10 keV, it was considered
as undetected. More than 60\% of energy depositions associated with neutrons were due to
single nuclear recoils. It is expected that events with multiple energy depositions in
different crystals will be rejected by the data acquisition system or off-line analysis.

\begin{table}[htbp]
\caption{Background event rate per year at 10-50 keV due to single nuclear
recoils from neutron
interactions in 103.68 kg of Ge. Neutron yield at production in the first column
is given in
neutrons/s/cm$^3$. }
\vspace{0.3cm}
\begin{center}
\begin{tabular}{|c|c|c|c|} \hline
Radioactive
& Inner Cu vessel
& External Cu vessel
& Plate  \\
contamination & \footnotesize{(mass=139 kg)} & \footnotesize{(mass=181 kg)}
& \footnotesize{(Cu mass = 22.1 kg or} \\
& & & \footnotesize{steel mass = 19.7 kg)} \\
\hline
1 ppb U in Cu  
& 1.49 & 1.35  &  0.680   \\
\footnotesize{(n yield = $1.24\times10^{-10}$)} & & & \\
\hline
1 ppb Th in Cu
& 0.130 & 0.114 & 0.0586   \\
\footnotesize{(n yield = $8.39\times10^{-12}$)} & & & \\
\hline
1 ppb U in stainless steel
& & &0.812  \\
\footnotesize{(n yield = $1.47\times10^{-10}$)} & & & \\
\hline
1 ppb Th in stainless steel
& & & 0.284  \\
(n yield = $5.16\times10^{-11}$) & & & \\
\hline

\end{tabular}
\label{table1}
\end{center}
\end{table}

To quantify the sensitivity of the experiment to \wi s 
and limitations due to
various backgrounds, we defined an energy range of 10-50 keV as the 
range for the data analysis from the future experiment.
Table \ref{table1} shows the expected event rate due to single nuclear recoils in the
10-50 keV energy range. Four different components were considered: inner copper vessel, 
outer copper vessel, copper plate, stainless steel plate. In the latter case, the copper plate
was substituted with a similar plate made of stainless steel. In all cases 1 ppb U and Th was
assumed. Stainless steel gives a slightly higher rate of events than copper due to 
higher cross-sections of \an reactions on some isotopes, assuming the same
concentrations of U and Th. It is known, however, that the usual concentrations of U and Th
in stainless steel are around 1 ppb whereas copper can be made 100 times less
contaminated with radioactive isotopes. Table \ref{table1} shows that about 20 kg of stainless
steel can produce a non-negligible neutron event rate in the detector.
This leads to the conclusion that no more than a few kg
of steel should be allowed to be used in the detector construction. The mass of copper and its
contamination should also be reduced to a minimum. A few thousand kg of copper
with a concentration of around 0.01 ppb of U and Th do not limit the sensitivity of the
detector made of 100 kg of Ge. Values given in Table \ref{table1} can be scaled up/down
for different masses of copper (stainless steel) and different concentrations of 
radioactive isotopes.
Scaling with the target mass or distance between copper and crystals is not straightforward.
To obtain accurate values, simulations should be repeated for different designs.

\section{Conclusions}

This work aimed mainly at presenting  
a tool for calculating neutron background associated with various radioactive isotope 
concentrations in different materials. This was done using the code \so, 
the libraries of which were extended by adding a large number of cross-sections for \an reactions 
and branching ratios. Cross-sections for \an reactions and branching ratios 
were evaluated using the code \emp. Neutron yields and spectra from uranium 
and thorium decay chains in different materials 
used in dark matter detectors and present in the rocks surrounding the 
experimental underground laboratories have been obtained. 
The radioactive contaminations of U and Th assumed for the calculations 
can easily be scaled to the actual values. 
If needed, spectra from other materials can be calculated using the same method.

We also showed first results concerning the estimated neutron background in a future 
large-scale dark matter experiment, based on cryogenic Ge detectors such as \eu. 
Based on the simulations carried out in present work, conclusions
can be derived that can help with the baseline design of the Ge experiment
expected to reach a sensitivity of about $10^{-10}$ pb to  \wi-nucleon cross-section:
\begin{enumerate}
\item A few tonnes of low radioactivity copper ($\le 0.01$ ppb U/Th) can be used in the
detector construction. This will produce a neutron background rate of less than
1 event per year at 10-50 keV and not limit detector sensitivity to  \wi s.
\item No more than a few kg of stainless steel or other material with concentrations of
about 1 ppb of U/Th are allowed inside the shielding.
\item Materials with higher concentrations of radioactive isotopes (more than
1 ppb of U/Th) should be avoided or their mass should be limited to less than a few
kg.

\end{enumerate}

\section{Acknowledgments}
This work has been supported by the ILIAS integrating activity
(Contract No. RII3-CT-2004-506222) as part of the EU FP6 programme 
in Astroparticle Physics. One of us (VT) would like to thank ILIAS
for the financial support of his PhD research.
We acknowledge also the support from 
the Science
and Technology Facility Council (UK).
We would like to thank the members of the EURECA Collaboration for fruitful
discussions.

\pagebreak

\pagebreak

\pagebreak

\pagebreak

\pagebreak

\pagebreak

\pagebreak

\pagebreak

 \end{document}